Revision 2

# Comparisons of the core and mantle compositions of Earth analogs from different terrestrial planet formation scenarios

Jesse T. Gu[1]*, Rebecca A. Fischer[1], Matthew C. Brennan[1], Matthew S. Clement[2], Seth A. Jacobson[3], Nathan A. Kaib[4], David P. O'Brien[5], and Sean N. Raymond[6]

[1]*Department of Earth and Planetary Sciences, Harvard University, Cambridge, MA, USA*

[2]*Earth and Planets Laboratory, Carnegie Institution for Science, Washington D.C., USA*

[3]*Department of Earth and Environmental Sciences, Michigan State University, East Lansing, MI, USA*

[4]*HL Dodge Department of Physics and Astronomy, University of Oklahoma, Norman, OK, USA*

[5]*Planetary Science Institute, Tucson, Arizona, USA*

[6]*Laboratoire d'Astrophysique de Bordeaux, Université de Bordeaux, CNRS, Bordeaux, France*

*Correspondence to jessegu@g.harvard.edu

**Abstract**

The chemical compositions of Earth's core and mantle provide insight into the processes that led to their formation. *N*-body simulations, on the other hand, generally do not contain chemical information, and seek to only reproduce the masses and orbits of the terrestrial planets. These simulations can be grouped into four potentially viable scenarios of Solar System formation (Classical, Annulus, Grand Tack, and Early Instability) for which we compile a total of 433 *N*-body simulations. We relate the outputs of these simulations to the chemistry of Earth's core and mantle using a melt-scaling law combined with a multi-stage model of core formation. We find the compositions of Earth analogs to be largely governed by the fraction of equilibrating embryo cores ($k_{core\_emb}$) and the initial embryo masses in *N*-body simulations, rather than the simulation type, where higher values of $k_{core\_emb}$ and larger initial embryo masses correspond to higher concentrations of Ni, Co, Mo, and W in Earth analog mantles and higher concentrations of Si and O in Earth analog cores. As a result, larger initial embryo masses require smaller values of $k_{core\_emb}$ to match Earth's mantle composition. On the other hand, compositions of Earth analog cores are sensitive to the temperatures of equilibration and $f$O$_2$ of accreting material. Simulation type may be important when considering magma ocean lifetimes, where Grand Tack simulations have the largest amounts of material accreted after the last giant impact. However, we cannot rule out any accretion scenarios or initial embryo masses due to the sensitivity of Earth's mantle composition to different parameters and the stochastic nature of *N*-body simulations. We use our compiled simulations to explore the relationship between initial embryo masses and the melting history of Earth analogs, where the complex interplay between the timing between impacts, magma ocean lifetimes, and volatile delivery could affect the compositions of Earth analogs formed from different simulation types. Comparing the last embryo impacts experienced by Earth analogs to specific Moon-forming scenarios, we find the characteristics of the Moon-forming impact are dependent on the initial conditions in *N*-body simulations where larger initial embryo masses promote larger and slower Moon-forming impactors. Mars-sized initial embryos are most consistent with the canonical hit-and-run scenario onto a solid mantle. Our results suggest that constraining the fraction of equilibrating impactor core ($k_{core}$) and the initial embryo masses in *N*-body simulations could be significant for understanding both Earth's accretion history and characteristics of the Moon-forming impact.

## 1. Introduction

The terrestrial planets formed in several stages on timescales on the order of 10–100 Myr (Righter and O'Brien, 2011). First, nebular gas condensed into dust, which accreted into planetesimals likely due to gravitational collapse triggered by the streaming instability (as reviewed in Johansen et al., 2014). This was followed by runaway growth, where the largest protoplanetary bodies grew the fastest either due to planetesimal accretion by pairwise growth (Kokubo and Ida, 1996) or pebble accretion (Lambrechts and Johansen, 2012; Levison et al., 2015). When the local surface density of planetesimals was consumed and/or the pebble flux has decreased, runaway growth transformed into oligarchic growth, creating a bimodal distribution of massive embryos and smaller planetesimals (Kokubo and Ida, 2000; Lambrechts et al., 2019). Finally, massive collisions between embryos (i.e., giant impacts) resulted in the present-day terrestrial planets (Chambers and Wetherill, 1998). *N*-body simulations of late-stage terrestrial planet formation seek to reproduce the orbits and masses of the terrestrial planets by calculating the gravitational interactions between a prescribed set of bodies that are representative of the final giant impact stage of terrestrial planet formation (Chambers, 2001). Classical simulations can be split into two types– Eccentric Jupiter and Saturn (EJS) simulations begin with Jupiter and Saturn in their modern orbits in eccentric orbits while Circular Jupiter and Saturn (CJS) simulations begin with Jupiter and Saturn closer than they are now, but in circular orbits (Fischer and Ciesla, 2014; O'Brien et al., 2006; Raymond et al., 2009; Woo et al., 2022). CJS simulations generally produce Mars analogs that are too large, whereas EJS simulations do slightly better in this regard but are self-inconsistent (see Raymond et al., 2009 for details). A number of other models have been proposed to address the small-Mars problem. Here, we focus on three well-studied scenarios which deplete mass in the Mars-forming region: 1) the "annulus" (or "low-mass asteroid belt") scenario, in which bodies are initially distributed in a narrow ring (Hansen, 2009; Kaib and Cowan, 2015; Raymond and Izidoro, 2017), 2) the Grand Tack scenario, in which Jupiter and Saturn migrate inward and then outward to their present locations (Walsh et al., 2011; Jacobson and Morbidelli, 2014; O'Brien et al., 2014), and 3) the Early Instability scenario, in which the gas giants undergo an orbital instability before Mars can grow (Clement et al., 2018, 2021; Liu et al., 2022). Earth analogs formed in each type of simulation are influenced by the dynamic evolution of the initial distribution of embryos and planetesimals. Different types of $N$-body simulations, which differ in terms of the growth rates and provenance of accreting bodies, will therefore form Earth analogs with unique accretion histories that accrete materials from distinct locations. In addition, different initial conditions prescribed between types of simulations, and even between simulations from the same suite, may result in different sizes and energies of accretionary impacts. Even though the probability of reproducing each terrestrial planet differs between simulation types, all simulation types have been shown to be plausibly capable of explaining the observed inner Solar System configuration (Raymond and Morbidelli, 2022).

Current state-of-the-art models of Earth's accretion and core formation integrate growth histories from $N$-body simulations with self-consistent evolution of oxygen fugacity ($fO_2$) to determine the partitioning of elements between metal and silicate following each impact (Fischer et al., 2017; Rubie et al., 2015). Experimental metal–silicate partitioning data are parameterized to capture the effects of pressure, temperature, composition, and $fO_2$ on the concentrations of each element in the resulting core and mantle (e.g., Fischer et al., 2015; Siebert et al., 2012). The aim of these models has been to reproduce Earth's observed mantle composition (McDonough and Sun, 1995; Palme and O'Neill, 2013) in terms of a set of major and minor elements, such as Mg, Ca, Al, Fe, Ni, Co, Nb, and Ta (Fischer et al., 2017; Rubie et al., 2011). However, previous studies have only investigated core formation under a single suite of $N$-body simulations, using Classical or Grand Tack simulations from a single study, which prevents comparison of different simulation types and initial conditions (Fischer et al., 2017; Rubie et al., 2015). Moreover, these studies used simplified metal–silicate equilibration parameters, where equilibration pressures ($P_{equil}$) were assumed to increase linearly such that they represented equilibration at a constant fraction of the growing core–mantle boundary pressure. While embryo masses generally increase as Earth accretes due to ongoing oligarchic growth elsewhere in the Solar System, the stochastic nature of $N$-body simulations means that the size and timing of impacts vary greatly between simulations. Comparing simulations from multiple studies can therefore constrain the range of possible accretional events Earth could have experienced during its growth. Other equilibration parameters, such as the fractions of terrestrial mantle and impactor core ($k_{core}$) that participate in metal–silicate equilibration, were set as a constant multiple of the impactor's mass and at a constant fraction, respectively. Simplifying these parameters ignores the amount of energy delivered to the proto-Earth by individual accretionary impacts and the volume of melting produced by each event (Abramov et al., 2012; Nakajima et al., 2021). In addition, it has been suggested that $k_{core}$ varies with impactor size, where larger impactor cores merge more efficiently with Earth's core and therefore experience lower degrees of equilibration (Marchi et al., 2018; Rubie et al., 2015).

de Vries et al. (2016) used the melt-scaling law of Abramov et al. (2012) to relate the energies of individual impacts to $P_{equil}$, assuming $P_{equil}$ occurred at the base of impact induced melt pools, over the course of accretion in Grand Tack simulations. They found $P_{equil}$ to depend on initial conditions of $N$-body simulations and the lifetime of magma oceans. However, the relationship between $P_{equil}$ and the resulting mantle compositions of Earth analogs remains unclear due to the absence of a core formation model. Furthermore, they only used Grand Tack simulations, yet it is possible that accretion histories of Earth analogs differ between different scenarios of Solar System formation. Here we build upon the study of de Vries et al. (2016) by compiling simulations from four different scenarios of Solar System formation and using a melt-scaling law based on hydrodynamic simulations (Nakajima et al., 2021) to determine $P_{equil}$ and the fraction of the proto-Earth's mantle melted by each impact ($f_{melt}$). Our results are integrated with state-of-the-art models of core formation (Fischer et al., 2017; Rubie et al., 2015) to explore the effects of varying simulation type and initial conditions within these different simulation types on the chemistry of Earth analogs formed. The compiled simulations and the results from our model are then used to compare the relationship between different $N$-body simulations, the initial conditions used within them, and volatile delivery and melting histories of Earth analogs.

## 2. Methods

### 2.1. *N*-body simulations

$N$-body simulations from different Solar System formation scenarios were compiled to use as inputs in our core formation model. A total of 48 Classical (CEJS) (O'Brien et al., 2006; Raymond et al., 2009), 110 Annulus (ANN) (Kaib and Cowan, 2015; Raymond and Izidoro, 2017), 142 Grand Tack (GT) (Jacobson and Morbidelli, 2014; O'Brien et al., 2014; Walsh et al., 2011), and 133 Early Instability (EI) (Clement et al., 2021) simulations were assembled to give a total of 433 simulations. Abbreviations of individual simulation names are given in Table 1 and will be used hereafter. Rows that contain simulation names in parentheses depict all simulations from a given study (e.g., CEJS-O06). Classical simulations from both studies were also split into CJS and EJS simulations to compare the two dynamical scenarios. GT simulations were split by both initial embryo to planetesimal mass ratio (e.g., GT 1:1) and initial embryo mass (e.g., GT-0.025) to compare different initial conditions. Earth analogs were defined as the body at the end of each simulation closest to 1 Earth mass and 1 AU within the ranges 0.8–1.25 Earth masses and 0.8–1.2 AU for a total of 109 Earth analogs. Such narrow ranges of masses and orbital radii were chosen to minimize the effects of planet mass and orbital radii of accreting material on Earth analog compositions (Fischer et al., 2017; Kaib and Cowan, 2015). We do not use the mass and orbit of Mars analogs as constraints because Mars' accretion history is unrelated to its final orbital parameters (Brennan et al., 2022). Table 1 also lists the initial conditions and number of Earth analogs produced from each suite of simulations. Initial embryo masses range from 0.005–0.48 Earth masses, with both the distributions and ranges of initial embryo masses varying between studies (Table 1). Most simulations used a bimodal distribution of larger embryos and smaller planetesimals, except for the ANN simulations from Kaib and Cowan (2015), in which simulations were run with equal-massed embryos only.

For each simulation that formed an Earth analog, the impact histories of bodies that eventually accreted to form the Earth analog were tracked. All bodies were assigned an initial composition based on relative non-volatile elemental abundances in CI chondrite. The abundances of refractory elements were enriched, and each body was equilibrated at a given $fO_2$ depending on its initial semi-major axis (Rubie et al., 2015). The initial $fO_2$ distribution followed a simple step-function with more reduced materials within the $fO_2$ step, as in previous studies (Fischer et al., 2017; Rubie et al., 2015). The $fO_2$ step was set to 2 AU with the outer $fO_2$ set to 1.5 log units below the iron-wüstite buffer ($\Delta$IW–1.5), consistent with Mars' accretion (Brennan et al., 2022), while the inner $fO_2$ was varied. Planetesimals were equilibrated at 0.1 GPa and 2000 K. To address differing initial embryo masses between simulation suites, we used a simple shell model to determine the initial pressure of equilibration in embryos. Here, we assumed a core mass fraction of 0.3, corresponding to equilibration of a CI chondrite composition without volatile elements at an $fO_2$ of $\sim\Delta IW$ $-3$. Even though the core mass fraction is dependent on the bulk composition and $fO_2$ of a differentiating body, we tested a case where embryos from beyond the $fO_2$ cutoff were equilibrated at a higher $P_{equil}$, corresponding to a smaller core mass fraction, resulting in differences in mantle composition, on average, of <1%. The total mass, density, and pressure of mantle layers were calculated from outside in until mass in the silicate reached 70% of the embryo's mass, resulting in a core–mantle boundary pressure. $P_{equil}$ was then set to be half of the core–mantle boundary pressure, which is consistent with the interpreted $P_{equil}$ on Mars (Brennan et al., 2020; Rai and van Westrenen, 2013; Righter and Chabot, 2011), since Mars may be a stranded embryo itself (Dauphas and Pourmand, 2011). The resulting parameterization for $P_{equil}$ of embryos was $P_{equil} = 112.06*M_{emb} + 0.37$ GPa, where $M_{emb}$ was the mass of the embryo in Earth masses.

## 2.2. Metal–silicate equilibration

For impactors >0.01 Earth masses (embryos), the impactor and target masses, impact velocity, and impact angle of each collision were taken from the outputs of each simulation and were used as inputs in the melt-scaling law of Nakajima et al. (2021) (Fig. 1). The melt-scaling law parameterizes outputs from hydrodynamic simulations to determine the volume and geometry of melting in the target's mantle, along with the pressure at the base of the melt pool. Impact angles were rounded to the nearest angle for which results are available (0°, 30°, 60°, or 90°) (e.g., for a 44° impact angle, results for 30° were used). We note impact angles in all simulations show a uniform distribution centered around 45 degrees (Supplementary Fig. 1) and that rounding in this way could result in additional uncertainties in the determined melt fraction. The surface entropy was set to 1100 J/K/kg, corresponding to surface temperatures of ~300 K. Assuming metal–silicate equilibration occurred at the base of the melt pool, $P_{equil}$ was set to the pressure at the base of the melt pool and $f_{melt}$ was set to the fraction of the mantle that was melted. We introduce a parameter, $k_{mantle\_melt}$, to describe the fraction of the melted mantle that equilibrates with the impactor's core. A schematic representation of embryo equilibration is shown in Fig. 1. For the sake of simplicity, we assumed instantaneous crystallization of magma oceans, but discuss the possible effects of longer magma ocean lifetimes in Section 4.1. Following each impact, the mantle was homogenized such that portions of the mantle that did not melt were assumed to fully mix with the melted portions. Embryo cores were also assumed to sink and merge with the proto-Earth's existing core before the next embryo impact and remained isolated from further equilibration. Even though the actual physics of metal–silicate mixing and equilibration during Earth's accretion are more complex (Deguen et al., 2014, 2011; Landeau et al., 2021), these simplifying assumptions allow us to relate impact energies to the conditions of core formation.

Smaller bodies <0.01 Earth masses (planetesimals) were below the threshold of masses compatible with the melt-scaling law of Nakajima et al. (2021) and were small enough that they may have been stranded in the proto-Earth's mantle following an impact (de Vries et al., 2016). The time it takes for a planetesimal's core to sink through the solid mantle and merge with Earth's core exceeds the time between embryo impacts such that planetesimals were assumed to equilibrate with the subsequent embryo impact (Fleck et al., 2018). Planetesimals that accreted after the last embryo impact were equilibrated at an assigned low pressure ($P_{equil\_ptsml}$) and target–to–impactor ratio of equilibrating silicate ($M_{melt\_ptsml}$) before the planetesimal's core merged with Earth's core and the equilibrated silicate mixed with Earth's mantle. Some embryos in CEJS simulations from Raymond et al. (2009) and ANN simulations from Kaib and Cowan (2015) were small enough that they were considered planetesimals during core formation (Table 1). Embryos in all other simulations were large enough that they were also considered embryos by the melt-scaling law. All embryos that eventually formed Earth analogs from CEJS simulations from Raymond et al. (2009) were >0.01 Earth masses, making this distinction only relevant for embryos from ANN simulations from Kaib and Cowan (2015). However, this discrepancy is only significant for small impactors that accrete after the last large impact, which our model is not very sensitive to (Table 2). We note that "planetesimals" in the context of core formation are all bodies <0.01 Earth masses, whereas it only apply to bodies <0.0025 Earth masses in the context of $N$-body simulations.

In contrast to $P_{equil}$ and $f_{melt}$, $k_{core}$ cannot be easily constrained from melt-scaling laws, especially for large embryo impacts. Instead, we assume constant reference values for $k_{core}$, but used different values for embryos ($k_{core\_emb} = 0.3$) and planetesimals ($k_{core\_ptsml} = 0.7$) to reflect the dependence of $k_{core}$ on the size of the impactor (Kendall and Melosh, 2016; Marchi et al., 2018). We note that $k_{core\_ptsml}$ is not set to 1 because it is possible that planetesimal cores remain in the proto-Earth's mantle and equilibrate during the next large impact. In this scenario, it is difficult to constrain the extent to which the original small impactor's core equilibrates. A compilation of equilibration parameters used, the ranges we tested, and the sensitivity of core and mantle compositions of Earth analogs are given in Table 2.

We followed the methodology detailed in Supplementary S2 of Rubie et al. (2011), as revised by Fischer et al. (2017) and Brennan et al. (2020) to evolve $f$O$_2$ self-consistently and calculate the composition of Earth's core and mantle following each equilibration event. The entire impactor mantle, along with a portion of the melted fraction of Earth's mantle ($f_{melt}*k_{mantle\_melt}$), participated in equilibration, where the fraction of the impactor's core that equilibrated was defined by $k_{core}$. In addition, material from planetesimals that impacted since the last embryo impact were added into the equilibrating mass. The core formation code of Brennan et al. (2020) was modified to incorporate impactor information from *N*-body simulations and was benchmarked against the results from Fischer et al. (2017). Parametrizations for $P_{equil}$ and $f_{melt}$ for each impactor were then incorporated according to the processes described above in place of the simple assumptions used in Fischer et al. (2017). The metal–silicate partitioning of elements were described by fits to experimental data as:

$$log_{10}(K_D^i) = a_i + \frac{b_i}{T} + \frac{c_i P}{T} \qquad (1)$$

for each element *i*, where *T* is the temperature in Kelvin, *P* is pressure in GPa, and $a_i$, $b_i$, and $c_i$ are fitting parameters. These parameters are detailed in Supplementary Table S1 and include the significant changes in partitioning at ~5 GPa for Si, O, Ni, and Co (Fischer et al., 2017 and references therein). We note that there may be more up-to-date partitioning parameterizations for Ni which have slightly different fitting coefficients (Huang and Badro, 2018). However, our results would not be affected significantly by incorporating these values. We also included fitting parameters for Nb, Ta, Mo, and W from Huang et al. (2020) and Huang et al. (2021) while Mg, Al, and Ca were assumed to be perfectly lithophile. $K_D$ is an exchange coefficient, defined in terms of partition coefficients (*D*), or in terms of the mole fractions (*X*) of elements and their oxides in the metal (*met*) and silicate (*sil*) as:

$$K_D^i = \frac{D_i}{D_{Fe}^{n/2}} = \frac{X_i^{met}/X_{iO_{n/2}}^{sil}}{(X_{Fe}^{met}/X_{FeO}^{sil})^{n/2}} \qquad (2)$$

where *n* is the valence of element *i*. Pressures of equilibration were determined for each impactor as described above, and the temperature of equilibration was described by a polynomial fit to the liquidos of Andrault et al. (2011). First, Fe Si, Ni, and O were partitioned, where the concentrations of Fe, Si, Ni, and O in the core-forming liquid were described as a function of the moles of FeO in the mantle. The moles of FeO in the mantle were then iterated until the moles of FeO in the mantle and the moles of Fe and O in the core-forming liquid were self-consistent. This was followed by the partitioning of the trace elements (Co, Nb, Ta, Mo, and W), where partitioning of the trace elements was iterated to ensure self-consistent description of molar abundances. The compositions of the proto-Earth's core and mantle were updated after each equilibration event, and these steps were repeated until the last embryo impact, after which planetesimals that accreted were equilibrated at a pressure defined by $P_{equil\_ptsml}$, with a portion of the mantle defined by $M_{melt\_ptsml}$, and a fraction of equilibrating core defined by $k_{core\_ptsml}$.

## 3. Results

### 3.1. Accretion histories and equilibration parameters

A comparison of the accretion histories from our different simulation suites shows the fast formation times of Earth analogs in GT and ANN simulations (Fig. 2, Supplementary Fig. S2). The time it takes for Earth analogs to reach 90% of their final mass ($t_{90}$) are given in Table 1. Despite large variations, these timescales are all consistent with the [182]W anomaly of Earth's mantle due to the large effect of varying $k_{core}$ on permissible timescales (Fischer and Nimmo, 2018). The distribution of impactor masses for each type of simulation are governed by initial embryo masses near 1 AU. As expected, simulations that begin with the largest embryos also have the largest median embryo masses regardless of simulation type (Fig. 2c–f). Furthermore, simulations containing large embryos also have more impacting embryos with mass >0.2 Earth masses. The impact velocities of embryo impacts are more consistent between different simulations and may be mainly controlled by the local gravitational environment at the time of the impact (Fig. 2g–j). Compared to other simulation types, impact velocities in GT and ANN-RI17 simulations are skewed towards smaller values due to dynamical friction from abundant unaccreted planetesimals, owing to their fast formation timescales. In contrast, ANN-KC15 simulations have fast formation times, don't have planetesimals, and as a result, are not affected by this dynamical friction.

Using the impactor masses, impact velocities, and impact angles of individual embryo impacts shown in Fig. 2 and Fig. S1 (Nakajima et al., 2021), $P_{equil}$ and $f_{melt}$ were determined for each impact and parameterized over the course of accretion for each Earth analog. Fig. 3 shows the distributions of $P_{equil}$ and $f_{melt}$ for all embryo impacts from each simulation suite. The combination of impactor masses and velocities shapes these distributions, such that simulations with larger impactor masses also have correspondingly higher $P_{equil}$ and $f_{melt}$. Larger initial embryo masses are therefore associated with more frequent high $P_{equil}$ and large mantle melting events ($f_{melt} > 0.8$). $P_{equil}$ averaged over all Earth analogs from each simulation type in 0.01 increments of mass fraction accreted are shown in Fig. 3i–j. These figures ignore the equilibration of planetesimals after the last embryo impact and are therefore representative of embryo collisions only. Incorporation of planetesimals would decrease the average $P_{equil}$ towards the end of accretion in all simulations depending on the fraction of material delivered after the last embryo impact. Average $P_{equil}$ are remarkably similar between different suites of simulations during the first 40% of accretion. EI simulations are an exception, owing to the large initial proto-Earth masses used in these simulations, which result in high values of average $P_{equil}$ due to our parameterization of embryo differentiation. Simulations with large initial embryo masses, which have more frequent large embryo collisions, are also able to reach the highest $P_{equil}$ on average. However, regardless of initial embryo mass, simulations with longer accretion timescales (CEJS and EI) reach higher maximum average $P_{equil}$ than those with shorter accretion timescales (ANN and GT). It is important to note that the curves shown in Fig. 3 are averaged over multiple Earth analogs and therefore do not fully capture the stochastic nature of the late stages of accretion, which results in large differences in $P_{equil}$ between Earth analogs, and even between Earth analogs from the same simulations suite. Here, GT simulations are split by their initial masses because there is no dependence on the initial embryo to planetesimal mass ratio (Supplementary Fig. S3).

### 3.2. Compositions of Earth analog cores and mantles

The core and mantle compositions of Earth analogs can be determined by combining the equilibration parameters and impact parameters determined from $N$-body simulations with our model of core formation. The metal–silicate partitioning of elements, as described by $K_D$, are sensitive to the partitioning of Fe between silicate and metal, or the $fO_2$ of equilibrating material (relative to the iron–wüstite (IW) buffer). For a constant set of equilibration parameters, average FeO concentrations vary greatly between Earth analogs from different simulations because of differences in the initial semi-major axis of accreting material (Fig. 4). Overall, Earth analogs produced by simulations from the same suite have similar mass-weighted average semi-major axes, regardless of initial embryo mass (Table 1, Supplementary Fig. S4). We adjust the $fO_2$ of accreting material by changing the inner $fO_2$ of the disk, such that on average, the FeO concentrations of Earth analogs formed match Earth's actual mantle FeO concentration (Supplementary Fig. S5). Adjusting the inner $fO_2$ to more reducing values has the same effect as moving the $fO_2$ step out in our model. On average, matching Earth's mantle FeO requires an inner $f$O$_2$ of ΔIW–3.16, –2.89, –2.59, and –2.60 for CEJS, GT, EI, and ANN simulations, respectively. The inner $f$O$_2$ for CEJS and GT simulations are slightly more oxidized than those used in Rubie et al. (2015) and Fischer et al. (2017), which used inner $f$O$_2$ values of IW–4 and –3.5, respectively. These differences can be attributed to the different simulations used in each study, where Fischer et al. (2017) used CEJS simulations from Fischer and Ciesla (2014), whereas Rubie et al. (2015) only used a subset of GT simulations from Jacobson and Morbidelli (2014).

Ni, Co, Nb, Ta, Mo, and W are siderophile elements that are either moderately refractory or refractory and have been used to trace core formation processes (Fischer et al., 2017; Huang et al., 2021, 2020; Jennings et al., 2021; Rubie et al., 2015). The partitioning behavior of each element is given in Supplementary Table S1 and their concentrations for each Earth analog formed are shown in Fig. 5, along with the concentrations of Si and O in Earth analog cores (Fig. 5 g–h). In Fig. 5, average $P_{equil}$ corresponds to $P_{equil}$ averaged over the growth history of each Earth analog within each simulation such that each point represents one Earth analog. In general, larger initial embryo masses result in higher NiO, CoO, MoO$_2$, and WO$_3$ in Earth analog mantles and higher Si and O in Earth analog cores. The concentrations of light elements in Earth's core remain debated, but we show core compositions from Hirose et al. (2021) for reference. Si and O combined make up ~2-7 wt% of the core, within the ranges allowed from geophysical and geochemical constraints (Fischer et al., 2014, 2011). Other light elements, such as H, C, and S, could also contribute to the density deficit of Earth's core (Blanchard et al., 2022; Fischer et al., 2020; Suer et al., 2017; Tagawa et al., 2021).

Some elements, such as Nb and Ta, match Earth's mantle (McDonough and Sun, 1995; Palme and O'Neill, 2013) for a wide range of average $P_{equil}$, whereas other elements, such as Ni,

Co, Mo, and W, are more sensitive to average $P_{equil}$. Fig. 6 shows the sensitivity of Earth analog core and mantle compositions to select parameters from Table 2. In addition to $k_{core\_emb}$, we find that the chosen initial conditions in $N$-body simulations play an important role in determining the composition of Earth's mantle. The changes resulting from $N$-body simulations are determined by taking difference between the average Earth analog compositions from ANN simulations from Kaib and Cowan (2015) (smallest embryos) and those from EI simulations (largest embryos). These differences result from the larger average $P_{equil}$ that correspond to large initial embryo masses (Fig. 3). The effects of $k_{core\_emb}$ on Ni and Co concentrations remedy the discrepancy in mantle composition between many Earth analogs and Earth. For example, larger values of $k_{core\_emb}$ would be required for Earth analogs with lower average $P_{equil}$. Mo and W, on the other hand, are systematically higher than Earth's mantle composition, even when considering the sensitivity of different parameters. It is possible that including the effects of C on Mo and W partitioning could make both elements more siderophile, reducing their mantle abundances (Jennings et al., 2021). In addition, the mantle compositions of Mo and W are highly uncertain (Liang et al., 2017). In contrast to mantle compositions, core compositions are more sensitive to certain core formation parameters (temperature for O and $f$O$_2$ for Si). The large deviations in core compositions indicate that Earth's core composition could vary significantly depending on the chosen conditions of core formation and could be more difficult to constrain. We emphasize that our goal is not to find the best set of parameters to match Earth's composition but show the compositions of Earth analogs as evidence that our model can reproduce Earth's mantle composition reasonably well, in addition to producing plausible core Si and O concentrations.

Another compositional effect not shown in Table 2 or Fig. 6 comes from planetesimals that accrete after the last embryo impact. A large fraction of this material must equilibrate, or else Earth's mantle siderophile element composition would greatly exceed mass estimates of the late veneer (Holzheid et al., 2000). Simulations with large percentages of material accreting after the last large impact have lower average $P_{equil}$ due to the equilibration of these planetesimals at low pressures. When looking specifically at GT simulations, it is difficult to distinguish any trend in composition with initial embryo mass due to larger initial embryo masses correlating with large percentages of material accreted after the last large impact (Fig. 5a–b and Table 1). In contrast, smaller percentages of material accreting after the last large impact in simulations with larger initial embryo to planetesimal mass ratios causes mantle NiO and CoO concentrations to increase (Table 1 and Supplementary Fig. S6). Nevertheless, the effects of material accreting after the last large impact, which are dependent on the type of $N$-body simulation, are not as significant as varying initial embryo masses.

## 4. Discussion

We have compiled $N$-body simulations and combined them with models of core formation in which $P_{equil}$ and $f_{melt}$ are parameterized using a melt-scaling law. We find Earth's mantle composition to be most sensitive to the initial embryo masses in $N$-body simulations and the chosen value of $k_{core}$. The sensitivity of Earth's mantle composition to these parameters allows Earth's mantle composition to be reproduced for different scenarios of Solar System formation and different initial conditions within these scenarios. Below, we explore the effects of assuming the crystallization timescale of magma oceans and potential implications of different accretion histories for the Moon-forming impact.

## 4.1. Magma ocean lifetimes and Earth's melting history

The results presented above assume instant magma ocean crystallization and planetesimals equilibrating with the next embryo impact. To test the effects of long-lived magma oceans, we use the opposite endmember scenario of infinite magma ocean lifetimes. Here all planetesimals equilibrate with magma oceans generated by the previous embryo impact. Planetesimals that accrete before the first embryo impact are equilibrated with the initial embryo that grew into the Earth analog. Following an embryo impact, the melt pool will isostatically adjust to form a global magma ocean on the order of $10^2$–$10^5$ years (Reese and Solomatov, 2006). Therefore, subsequent planetesimals will equilibrate at the base of the global magma ocean rather than with the melt pool. In contrast to the instant crystallization scenario, surface entropy in the long-lived magma ocean case was set to 3160 J/K/kg, corresponding to surface temperatures of ~2000 K. Table 2 and Fig. 6 show the sensitivities of Earth's core and mantle compositions to variations in magma ocean lifetime. Supplementary Fig. S7 shows the differences (long-lived magma oceans – instant crystallization) in average $P_{equil}$ and mantle NiO and CoO concentrations for each simulation that produced an Earth analog. In general, average $P_{equil}$ is not shifted significantly, except in Grand Tack simulations, which have large percentages of material accreted after the last giant impact, due to the equilibration of planetesimals at the base of a deep magma ocean formed by the previous embryo impact. For other simulation types, these effects are overshadowed by $k_{core\_emb}$ and initial embryo mass for mantle composition but can be significant for the O concentration in the core, due to the large effect of temperature on O partitioning.

Compared to the two endmember scenarios we explored, realistic magma ocean lifetimes depend on the timing of embryo impacts and the efficiency of heat loss from the proto-Earth's interior to space. The presence or absence of an atmosphere, which hinges on the complex interplay between volatile delivery and atmosphere erosion, would therefore strongly influence the equilibration of planetesimals during accretion and could affect the compositions of Earth analogs formed in certain simulation types (Elkins-Tanton, 2008; Lebrun et al., 2013; Sakuraba et al., 2021). Assuming persistent atmospheres throughout accretion, the different timescales of Earth's accretion between different simulation types could be related to varying proportions of magma oceans that persist until the following embryo impact (Fig. 7). Specifically, fast accretion timescales in Grand Tack and Annulus result in 70.1% and 76.9% of embryos impacting within 2 Myrs of each other. These fast accretion timescales could promote persistent magma oceans and planetesimal impacts onto existing magma oceans (de Vries et al., 2016). In contrast, CEJS and EI simulations have less frequent embryo impacts, with only 30.0% and 5.3% of impacts occurring within <2Myrs apart. Despite these correlations, the stochastic nature of $N$-body simulations complicates the interpretation of these data. For example, small amounts of mass originating from large semi-major axes could contribute a significant quantity of Earth's volatiles at specific times during accretion. Furthermore, magma ocean lifetimes are dependent on the mass and composition of existing atmospheres, which could be related to the delivery of different volatile species and differences in their solubilities (Gaillard et al., 2022; Lichtenberg et al., 2021). Whether the proto-Earth could sustain an atmosphere during the giant impact stage of accretion and its relationship to the composition of Earth's core and mantle still needs to be explored.

The masses assigned to embryos at the start of the giant impact stage of accretion define the initial conditions of *N*-body simulations. Jacobson and Morbidelli (2014) predicted that Mars-sized embryos would best match the Solar System's architecture. It is often assumed in *N*-body simulations that all embryos begin with equal masses. Recent advances in simulating the formation of embryos suggests that the presence of a dissipating gas disk promotes formation of the largest embryos inside 1 AU, with the largest embryos reaching up to ~10–50% of Earth's mass (Clement et al., 2020; Walsh and Levison, 2019; Woo et al., 2021). Our results suggest that all initial conditions can match Earth's mantle composition due to the unconstrained value of $k_{core}$. However, the number of embryo impacts is independent of simulation type and decreases with the average initial embryo mass from which Earth analogs form (Fig. 8). Within GT simulations, increasing the initial embryo to planetesimal mass ratio increases the number of embryo impacts, because planetesimals supply less mass during Earth's formation. For CEJS and EI simulations, where magma oceans are more likely to crystallize before the next embryo impact, the number of embryo impacts is also more likely to be representative of the number of magma oceans experienced by Earth analogs. For GT and ANN simulations, which are more likely to have persistent magma oceans, the number of embryo impacts corresponds to the maximum number of magma oceans Earth analogs would have had. Earth analogs in EI simulations, which use initial conditions from the outputs of an embryo growth model with ~1:1 embryo to planetesimal mass ratios, likely experienced between 2–5 magma ocean events. Geochemical estimates suggest that Earth experienced at least two magma oceans during its accretion (Tucker and Mukhopadhyay, 2014). We show how the initial embryo masses and embryo to planetesimal mass ratios can be used to place constraints on the maximum number of magma oceans and outgassing events Earth experienced. Future constraints on the geochemical consequence of magma oceans may therefore place constraints on the masses of embryos in the early Solar System.

## 4.2. Implications for Moon formation

The likelihood of specific Moon-forming impact scenarios and the resulting melting of Earth's mantle can be evaluated by focusing on the last embryo impact, or sequence of embryo impacts, experienced by each Earth analog (Fig. 9) (Jacobson and Morbidelli, 2014). Increasing initial embryo masses results in larger last embryo impacts that melt a large fraction (>90%) of Earth's mantle ($f_{melt}$). Equal-massed embryos of 0.005 Earth masses used in ANN-KC15 simulations result in Moon-forming impactors that are too small to match any Moon-forming scenario. The most probable Moon-forming scenarios based on impactor masses are the canonical hit-and-run and rapidly rotating Earth scenarios (Canup and Asphaug, 2001; Ćuk and Stewart, 2012; Reufer et al., 2012). Simulations that begin with Mars-sized embryos are most consistent with the canonical hit-and-run scenario, whereas those with smaller embryos have masses most consistent with the rapidly rotating Earth scenario. On the other hand, equal size impactors are rarely achieved, although the probability of such a scenario could be increased with larger initial embryos, or during pebble accretion (Canup, 2012; Johansen et al., 2021). By also considering impact velocities, it becomes difficult to simultaneously match the scaled impactor masses and high impact velocities required by the rapidly rotating Earth scenario (Kaib and Cowan, 2015). However, we do note that smaller initial embryo masses correspond to higher likelihoods of fast ($v_{rel} > 2$) last embryo impacts (Fig. 9e–g). Our results thus suggest that the probability of each Moon-forming scenario is dependent on the initial conditions in $N$-body simulations, where larger initial embryo masses promote larger and slower impactors. Mars-sized initial embryos are most consistent with the canonical hit-and-run scenario. Constraining the initial conditions in *N*-body simulations will therefore aid in understanding the likelihood of last embryo impacts that fall within the range allowed by each Moon-forming scenario.

Recent theories of Moon formation have emerged that add to the range of possible scenarios presented above. A canonical impact onto an existing magma ocean aids in matching the compositional similarities between the Earth and the Moon (Hosono et al., 2019). Even though impactor masses and impact velocities allowed in such a scenario are similar to the canonical hit-and-run, the probability that the last embryo impact occurs onto an existing magma ocean depends on the time between the last two embryo impacts. We find this probability to be <5% for CEJS and EI simulations, and <15% for GT and ANN simulations, assuming a maximum magma ocean lifetime of 2 Myrs (Supplementary Fig. S8). When focusing only on GT simulations, this probability increases to 22.5%. It is also a possibility that the Moon formed from a series of impacts throughout Earth's accretion (Rufu et al., 2017). Interestingly, 99 out of 109 (90.8%) of Earth analogs experience complete mantle melting at some point during accretion. Therefore, Earth analogs that don't experience large Moon-forming impacts are still likely to have experienced complete mantle melting from a prior large embryo impact. Such impacts could aid in the formation of moonlets. Evaluating the likelihood of Moon formation from multiple impacts is beyond the scope of the current work but should be investigated by future studies that incorporate realistic impact histories from *N*-body simulations with hydrodynamic impact simulations.

## 5. Conclusions

We have compiled *N*-body simulations covering four models of Solar System formation. Building upon previous models of accretion and core formation (Fischer et al., 2017; Rubie et al., 2011), we incorporate the melt-scaling law of Nakajima et al. (2021) to explore the relationships between the compiled *N*-body simulations, equilibration parameters, and Earth's mantle composition. We find Earth's mantle composition to be most sensitive to the initial embryo mass in *N*-body simulations and the chosen value of $k_{core\_emb}$. These sensitivities allow Earth's mantle composition to be matched for different scenarios of Solar System formation and initial conditions within. Larger initial embryo masses require smaller values of $k_{core}$, on average, to match Earth's mantle composition. Considering interactions between accretion timescales, magma oceans, volatile delivery, Earth's mantle composition may be affected by magma ocean lifetimes depending on the time between embryo impacts. Characteristics of last embryo impacts suggest that Mars-sized embryos are most consistent with the canonical hit-and-run scenario onto a solid mantle. However, future constraints on the initial embryo masses in *N*-body simulations and the values of $k_{core\_emb}$ could yield further insights into Earth's accretion history and the Moon-forming impact.

## 6.  Acknowledgements

This work was supported in part by National Science Foundation Graduate Research Fellowships awarded to J.T. Gu and M.C. Brennan (DGE1745303), a NASA Emerging Worlds grant (80NSSC21K0388) and an NSF grant (EAR-2054912) both awarded to R.A. Fischer, and an NSF CAREER Award to N.A Kaib (1846388). We thank M. Nakajima for providing the publicly available melt-scaling law and assistance with its use in our model. We also thank J. Dong and H. Fu for helpful discussions and advice throughout the duration of this project.

Finally, we thank D. Rubie and an anonymous reviewer for their suggestions that have substantially improved the manuscript.

**References**

Abramov, O., Wong, S.M., Kring, D.A., 2012. Differential melt scaling for oblique impacts on terrestrial planets. Icarus 218, 906–916. https://doi.org/10.1016/j.icarus.2011.12.022

Andrault, D., Bolfan-Casanova, N., Nigro, G. lo, Bouhifd, M.A., Garbarino, G., Mezouar, M., 2011. Solidus and liquidus profiles of chondritic mantle: Implication for melting of the Earth across its history. Earth Planet Sci Lett. https://doi.org/10.1016/j.epsl.2011.02.006

Blanchard, I., Rubie, D.C., Jennings, E.S., Franchi, I.A., Zhao, X., Petitgirard, S., Miyajima, N., Jacobson, S.A., Morbidelli, A., 2022. The metal–silicate partitioning of carbon during Earth's accretion and its distribution in the early solar system. Earth Planet Sci Lett 580. https://doi.org/10.1016/j.epsl.2022.117374

Brennan, M.C., Fischer, R.A., Irving, J.C.E., 2020. Core formation and geophysical properties of Mars. Earth Planet Sci Lett 530, 115923. https://doi.org/10.1016/j.epsl.2019.115923

Brennan, M.C., Fischer, R.A., Nimmo, F., O'Brien, D.P., 2022. Timing of Martian core formation from models of Hf–W evolution coupled with N-body simulations. Geochim Cosmochim Acta 316, 295–308. https://doi.org/10.1016/j.gca.2021.09.022

Canup, R.M., 2012. Forming a moon with an Earth-like composition via a giant impact. Science (1979) 338. https://doi.org/10.1126/science.1226073

Canup, R.M., Asphaug, E., 2001. Origin of the Moon in a giantimpact near the end of theEarth's formation. Nature 412, 708–712.

Chambers, J.E., 2001. Making More Terrestrial Planets. Icarus 152. https://doi.org/10.1006/icar.2001.6639

Chambers, J.E., Wetherill, G.W., 1998. Making the Terrestrial Planets: N-Body Integrations of Planetary Embryos in Three Dimensions. Icarus 136. https://doi.org/10.1006/icar.1998.6007

Clement, M.S., Kaib, N.A., Chambers, J.E., 2020. Embryo Formation with GPU Acceleration: Reevaluating the Initial Conditions for Terrestrial Accretion. Planet Sci J 1, 18. https://doi.org/10.3847/psj/ab91aa

Clement, M.S., Kaib, N.A., Raymond, S.N., Chambers, J.E., 2021. The early instability scenario: Mars' mass explained by Jupiter's orbit. Icarus 367, 114585. https://doi.org/10.1016/j.icarus.2021.114585

Clement, M.S., Kaib, N.A., Raymond, S.N., Walsh, K.J., 2018. Mars' growth stunted by an early giant planet instability. Icarus 311, 340–356. https://doi.org/10.1016/j.icarus.2018.04.008

Ćuk, M., Stewart, S.T., 2012. Making the moon from a fast-spinning earth: A giant impact followed by resonant despinning. Science (1979) 338, 1047–1052. https://doi.org/10.1126/science.1225542

Dauphas, N., Pourmand, A., 2011. Hf-W-Th evidence for rapid growth of Mars and its status as a planetary embryo. Nature. https://doi.org/10.1038/nature10077

de Vries, J., Nimmo, F., Melosh, H.J., Jacobson, S.A., Morbidelli, A., Rubie, D.C., 2016. Impact-induced melting during accretion of the Earth. Prog Earth Planet Sci 3. https://doi.org/10.1186/s40645-016-0083-8

Deguen, R., Landeau, M., Olson, P., 2014. Turbulent metal-silicate mixing, fragmentation, and equilibration in magma oceans. Earth Planet Sci Lett. https://doi.org/10.1016/j.epsl.2014.02.007

Deguen, R., Olson, P., Cardin, P., 2011. Experiments on turbulent metal-silicate mixing in a magma ocean. Earth Planet Sci Lett. https://doi.org/10.1016/j.epsl.2011.08.041

Elkins-Tanton, L.T., 2008. Linked magma ocean solidification and atmospheric growth for Earth and Mars. Earth Planet Sci Lett 271, 181–191. https://doi.org/10.1016/j.epsl.2008.03.062

Fischer, R.A., Campbell, A.J., Caracas, R., Reaman, D.M., Heinz, D.L., Dera, P., Prakapenka, V.B., 2014. Equations of state in the Fe-FeSi system at high pressures and temperatures. J Geophys Res Solid Earth. https://doi.org/10.1002/2013JB010898

Fischer, R.A., Campbell, A.J., Ciesla, F.J., 2017. Sensitivities of Earth's core and mantle compositions to accretion and differentiation processes. Earth Planet Sci Lett 458, 252–262. https://doi.org/10.1016/j.epsl.2016.10.025

Fischer, R.A., Campbell, A.J., Shofner, G.A., Lord, O.T., Dera, P., Prakapenka, V.B., 2011. Equation of state and phase diagram of FeO. Earth Planet Sci Lett. https://doi.org/10.1016/j.epsl.2011.02.025

Fischer, R.A., Ciesla, F.J., 2014. Dynamics of the terrestrial planets from a large number of N-body simulations. Earth Planet Sci Lett 392, 28–38. https://doi.org/10.1016/j.epsl.2014.02.011

Fischer, R.A., Cottrell, E., Hauri, E., Lee, K.K.M., le Voyer, M., 2020. The carbon content of Earth and its core. Proceedings of the National Academy of Sciences 117, 8743–8749. https://doi.org/10.1073/pnas.1919930117

Fischer, R.A., Nakajima, Y., Campbell, A.J., Frost, D.J., Harries, D., Langenhorst, F., Miyajima, N., Pollok, K., Rubie, D.C., 2015. High pressure metal–silicate partitioning of Ni, Co, V, Cr, Si, and O. Geochim Cosmochim Acta 167, 177–194. https://doi.org/10.1016/j.gca.2015.06.026

Fischer, R.A., Nimmo, F., 2018. Effects of core formation on the Hf–W isotopic composition of the Earth and dating of the Moon-forming impact. Earth Planet Sci Lett. https://doi.org/10.1016/j.epsl.2018.07.030

Fleck, J.R., Rains, C.L., Weeraratne, D.S., Nguyen, C.T., Brand, D.M., Klein, S.M., McGehee, J.M., Rincon, J.M., Martinez, C., Olson, P.L., 2018. Iron diapirs entrain silicates to the core and initiate thermochemical plumes. Nat Commun 9, 1–12. https://doi.org/10.1038/s41467-017-02503-2

Gaillard, F., Bernadou, F., Roskosz, M., Bouhifd, M.A., Marrocchi, Y., Iacono-Marziano, G., Moreira, M., Scaillet, B., Rogerie, G., 2022. Redox controls during magma ocean degassing. Earth Planet Sci Lett 577. https://doi.org/10.1016/j.epsl.2021.117255

Hansen, B.M.S., 2009. Formation of the terrestrial planets from a narrow annulus. Astrophysical Journal 703, 1131–1140. https://doi.org/10.1088/0004-637X/703/1/1131

Hirose, K., Wood, B., Vočadlo, L., 2021. Light elements in the Earth's core. Nat Rev Earth Environ. https://doi.org/10.1038/s43017-021-00203-6

Holzheid, A., Sylvester, P., O'Neill, H.S.C., Ruble, D.C., Palme, H., 2000. Evidence for a late chondritic veneer in the Earth's mantle from high-pressure partitioning of palladium and platinum. Nature 406. https://doi.org/10.1038/35019050

Hosono, N., Karato, S. ichiro, Makino, J., Saitoh, T.R., 2019. Terrestrial magma ocean origin of the Moon. Nat Geosci 12, 418–423. https://doi.org/10.1038/s41561-019-0354-2

Huang, D., Badro, J., Siebert, J., 2020. The niobium and tantalum concentration in the mantle constrains the composition of Earth's primordial magma ocean. Proc Natl Acad Sci U S A 117, 27893–27898. https://doi.org/10.1073/pnas.2007982117

Huang, D., Siebert, J., Badro, J., 2021. High pressure partitioning behavior of Mo and W and late sulfur delivery during Earth's core formation. Geochim Cosmochim Acta 310. https://doi.org/10.1016/j.gca.2021.06.031

Jacobson, S.A., Morbidelli, A., 2014. Lunar and terrestrial planet formation in the Grand Tack scenario. Philosophical Transactions of the Royal Society A: Mathematical, Physical and Engineering Sciences. https://doi.org/10.1098/rsta.2013.0174

Jennings, E.S., Jacobson, S.A., Rubie, D.C., Nakajima, Y., Vogel, A.K., Rose-Weston, L.A., Frost, D.J., 2021. Metal–silicate partitioning of W and Mo and the role of carbon in controlling their abundances in the bulk silicate earth. Geochim Cosmochim Acta 293. https://doi.org/10.1016/j.gca.2020.09.035

Johansen, A., Blum, J., Tanaka, H., Ormel, C., Bizzarro, M., Rickman, H., 2014. The Multifaceted Planetesimal Formation Process, in: Protostars and Planets VI. https://doi.org/10.2458/azu_uapress_9780816531240-ch024

Johansen, A., Ronnet, T., Bizzarro, M., Schiller, M., Lambrechts, M., Nordlund, A., Lammer, H., 2021. A pebble accretion model for the formation of the terrestrial planets in the solar system. Sci Adv 7. https://doi.org/10.1126/sciadv.abc0444

Kaib, N.A., Cowan, N.B., 2015. The feeding zones of terrestrial planets and insights into Moon formation. Icarus 252, 161–174. https://doi.org/10.1016/j.icarus.2015.01.013

Kendall, J.D., Melosh, H.J., 2016. Differentiated planetesimal impacts into a terrestrial magma ocean: Fate of the iron core. Earth Planet Sci Lett 448, 24–33. https://doi.org/10.1016/j.epsl.2016.05.012

Kokubo, E., Ida, S., 2000. Formation of Protoplanets from Planetesimals in the Solar Nebula. Icarus 143. https://doi.org/10.1006/icar.1999.6237

Kokubo, E., Ida, S., 1996. On runaway growth of planetesimals. Icarus 123. https://doi.org/10.1006/icar.1996.0148

Lambrechts, M., Johansen, A., 2012. Rapid growth of gas-giant cores by pebble accretion. Astron Astrophys 544. https://doi.org/10.1051/0004-6361/201219127

Lambrechts, M., Morbidelli, A., Jacobson, S.A., Johansen, A., Bitsch, B., Izidoro, A., Raymond, S.N., 2019. Formation of planetary systems by pebble accretion and migration: How the radial pebble flux determines a terrestrial-planet or super-Earth growth mode. Astron Astrophys 627. https://doi.org/10.1051/0004-6361/201834229

Landeau, M., Deguen, R., Phillips, D., Neufeld, J.A., Lherm, V., Dalziel, S.B., 2021. Metal-silicate mixing by large Earth-forming impacts. Earth Planet Sci Lett 564, 116888. https://doi.org/10.1016/j.epsl.2021.116888

Lebrun, T., Massol, H., Chassefière, E., Davaille, A., Marcq, E., Sarda, P., Leblanc, F., Brandeis, G., 2013. Thermal evolution of an early magma ocean in interaction with the atmosphere. Journal of Geophysical Research E: Planets 118. https://doi.org/10.1002/jgre.20068

Levison, H.F., Kretke, K.A., Duncan, M.J., 2015. Growing the gas-giant planets by the gradual accumulation of pebbles. Nature 524. https://doi.org/10.1038/nature14675

Liang, Y.H., Halliday, A.N., Siebert, C., Fitton, J.G., Burton, K.W., Wang, K.L., Harvey, J., 2017. Molybdenum isotope fractionation in the mantle. Geochim Cosmochim Acta 199. https://doi.org/10.1016/j.gca.2016.11.023

Lichtenberg, T., Bower, D.J., Hammond, M., Boukrouche, R., Sanan, P., Tsai, S.M., Pierrehumbert, R.T., 2021. Vertically Resolved Magma Ocean–Protoatmosphere Evolution: H2, H2O, CO2, CH4, CO, O2, and N2 as Primary Absorbers. J Geophys Res Planets 126. https://doi.org/10.1029/2020JE006711

Liu, B., Raymond, S.N., Jacobson, S.A., 2022. Early Solar System instability triggered by dispersal of the gaseous disk. Nature 604, 643–646. https://doi.org/10.1038/s41586-022-04535-1

Marchi, S., Canup, R.M., Walker, R.J., 2018. Heterogeneous delivery of silicate and metal to the Earth by large planetesimals. Nat Geosci 11, 77–81. https://doi.org/10.1038/s41561-017-0022-3

McDonough, W.F.F., Sun, S. -s. s., 1995. The composition of the Earth. Chem Geol 120, 223–253. https://doi.org/10.1016/0009-2541(94)00140-4

Nakajima, M., Golabek, G.J., Wünnemann, K., Rubie, D.C., Burger, C., Melosh, H.J., Jacobson, S.A., Manske, L., Hull, S.D., 2021. Scaling laws for the geometry of an impact-induced magma ocean. Earth Planet Sci Lett 568, 116983. https://doi.org/10.1016/j.epsl.2021.116983

O'Brien, D.P., Morbidelli, A., Levison, H.F., 2006. Terrestrial planet formation with strong dynamical friction. Icarus 184, 39–58. https://doi.org/10.1016/j.icarus.2006.04.005

O'Brien, D.P., Walsh, K.J., Morbidelli, A., Raymond, S.N., Mandell, A.M., 2014. Water delivery and giant impacts in the "Grand Tack" scenario. Icarus 239, 74–84. https://doi.org/10.1016/j.icarus.2014.05.009

Palme, H., O'Neill, H., 2013. Cosmochemical Estimates of Mantle Composition, 2nd ed, Treatise on Geochemistry: Second Edition. Elsevier Ltd. https://doi.org/10.1016/B978-0-08-095975-7.00201-1

Rai, N., van Westrenen, W., 2013. Core-mantle differentiation in Mars. J Geophys Res Planets 118. https://doi.org/10.1002/jgre.20093

Raymond, S.N., Izidoro, A., 2017. The empty primordial asteroid belt. Sci Adv 3. https://doi.org/10.1126/sciadv.1701138

Raymond, S.N., Morbidelli, A., 2022. Planet Formation: Key Mechanisms and Global Models. https://doi.org/10.1007/978-3-030-88124-5_1

Raymond, S.N., O'Brien, D.P., Morbidelli, A., Kaib, N.A., 2009. Building the terrestrial planets: Constrained accretion in the inner Solar System. Icarus 203, 644–662. https://doi.org/10.1016/j.icarus.2009.05.016

Reese, C.C., Solomatov, V.S., 2006. Fluid dynamics of local martian magma oceans. Icarus 184, 102–120. https://doi.org/10.1016/j.icarus.2006.04.008

Reufer, A., Meier, M.M.M., Benz, W., Wieler, R., 2012. A hit-and-run giant impact scenario. Icarus 221. https://doi.org/10.1016/j.icarus.2012.07.021

Righter, K., Chabot, N.L., 2011. Moderately and slightly siderophile element constraints on the depth and extent of melting in early Mars. Meteorit Planet Sci 46. https://doi.org/10.1111/j.1945-5100.2010.01140.x

Righter, K., O'Brien, D.P., 2011. Terrestrial planet formation. Proc Natl Acad Sci U S A. https://doi.org/10.1073/pnas.1013480108

Rubie, D.C., Frost, D.J., Mann, U., Asahara, Y., Nimmo, F., Tsuno, K., Kegler, P., Holzheid, A., Palme, H., 2011. Heterogeneous accretion, composition and core–mantle differentiation of the Earth. Earth Planet Sci Lett 301, 31–42. https://doi.org/10.1016/j.epsl.2010.11.030

Rubie, D.C., Jacobson, S.A., Morbidelli, A., O'Brien, D.P., Young, E.D., de Vries, J., Nimmo, F., Palme, H., Frost, D.J., 2015. Accretion and differentiation of the terrestrial planets with implications for the compositions of early-formed Solar System bodies and accretion of water. Icarus 248, 89–108. https://doi.org/10.1016/j.icarus.2014.10.015

Rufu, R., Aharonson, O., Perets, H.B., 2017. A multiple-impact origin for the Moon. Nat Geosci 10, 89–94. https://doi.org/10.1038/ngeo2866

Sakuraba, H., Kurokawa, H., Genda, H., Ohta, K., 2021. Numerous chondritic impactors and oxidized magma ocean set Earth's volatile depletion. Sci Rep 11. https://doi.org/10.1038/s41598-021-99240-w

Siebert, J., Badro, J., Antonangeli, D., Ryerson, F.J., 2012. Metal–silicate partitioning of Ni and Co in a deep magma ocean. Earth Planet Sci Lett 321–322, 189–197. https://doi.org/10.1016/j.epsl.2012.01.013

Suer, T.-A., Siebert, J., Remusat, L., Menguy, N., Fiquet, G., 2017. A sulfur-poor terrestrial core inferred from metal–silicate partitioning experiments. Earth Planet Sci Lett 469, 84–97. https://doi.org/10.1016/j.epsl.2017.04.016

Tagawa, S., Sakamoto, N., Hirose, K., Yokoo, S., Hernlund, J., Ohishi, Y., Yurimoto, H., 2021. Experimental evidence for hydrogen incorporation into Earth's core. Nat Commun 12. https://doi.org/10.1038/s41467-021-22035-0

Tucker, J.M., Mukhopadhyay, S., 2014. Evidence for multiple magma ocean outgassing and atmospheric loss episodes from mantle noble gases. Earth Planet Sci Lett. https://doi.org/10.1016/j.epsl.2014.02.050

Walsh, K.J., Levison, H.F., 2019. Planetesimals to terrestrial planets: Collisional evolution amidst a dissipating gas disk. Icarus 329, 88–100. https://doi.org/10.1016/j.icarus.2019.03.031

Walsh, K.J., Morbidelli, A., Raymond, S.N., O'Brien, D.P., Mandell, A.M., 2011. A low mass for Mars from Jupiter's early gas-driven migration. Nature. https://doi.org/10.1038/nature10201

Woo, J.M.Y., Brasser, R., Grimm, S.L., Timpe, M.L., Stadel, J., 2022. The terrestrial planet formation paradox inferred from high-resolution N-body simulations. Icarus 371. https://doi.org/10.1016/j.icarus.2021.114692

Woo, J.M.Y., Grimm, S., Brasser, R., Stadel, J., 2021. Growing Mars fast: High-resolution GPU simulations of embryo formation. Icarus 359. https://doi.org/10.1016/j.icarus.2021.114305

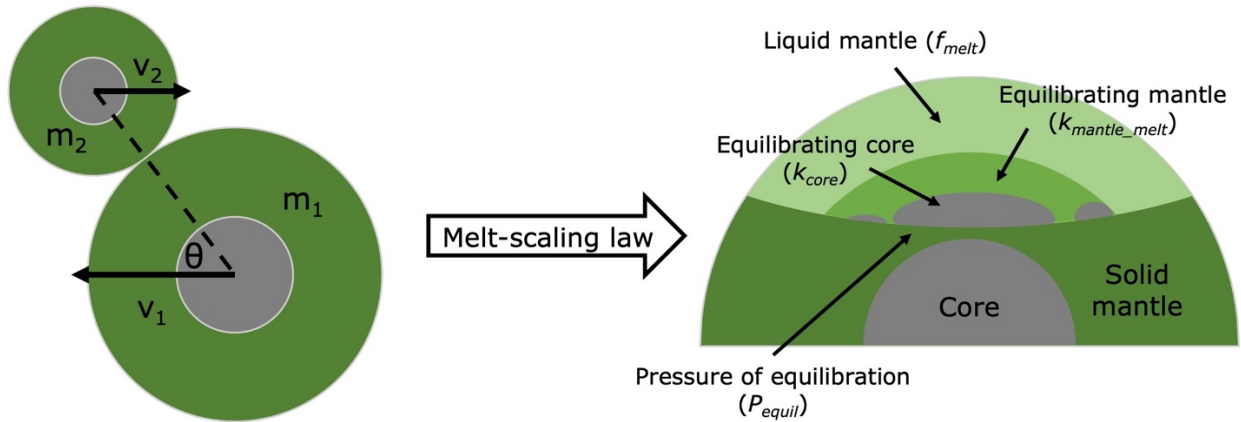

**Fig. 1.** Schematic representation of the melt-scaling law from Nakajima et al. (2021) and equilibration of embryos at the base of the melt pool formed from an embryo impact. Masses ($m_1$ and $m_2$), velocities ($v_1$ and $v_2$), and impact angle ($\theta$) are labeled on the left. $P_{equil}$ is the pressure at the base of the melt pool, $f_{melt}$ is the fraction of the target's mantle that is melted, $k_{mantle\_melt}$ is the fraction of the melted mantle that equilibrates with the impactor's core (such that $f_{melt}*k_{mantle\_melt}$ = fraction of the whole mantle that equilibrates), and $k_{core}$ is the fraction of the impactor's core that equilibrates.

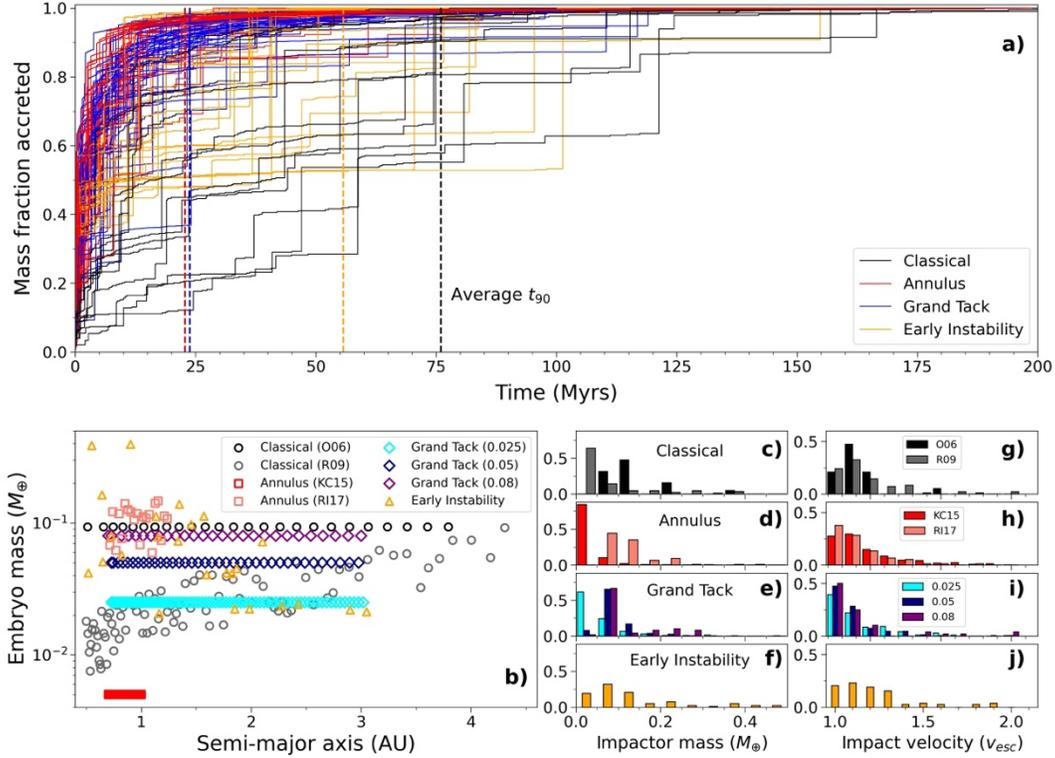

**Fig. 2.** Properties of Earth's accretion history from different *N*-body simulations and scenarios of Solar System formation. a) Accretion histories of Earth analogs where dashed vertical lines represent the average time at which Earth analogs from each scenario reach 90% of their final mass ($t_{90}$). b) Distributions of initial embryo masses used as initial conditions in *N*-body simulations. The symbols on the figure represent initial conditions for only one simulation from each suite. c–f) Normalized histograms of impactor masses and g–j) impact velocities of embryo collisions onto Earth analogs. Impactor masses and impact velocities are binned in increments of 0.05 Earth masses and 0.1, respectively. Abbreviations are as follows: O06 (O'Brien et al., 2006), R09 (Raymond et al., 2009), KC15 (Kaib and Cowan, 2015), RI17 (Raymond and Izidoro, 2017a), and 0.025, 0.05, and 0.08 represent the initial embryo masses, in Earth masses, of Grand Tack simulations from Jacobson and Morbidelli (2014).

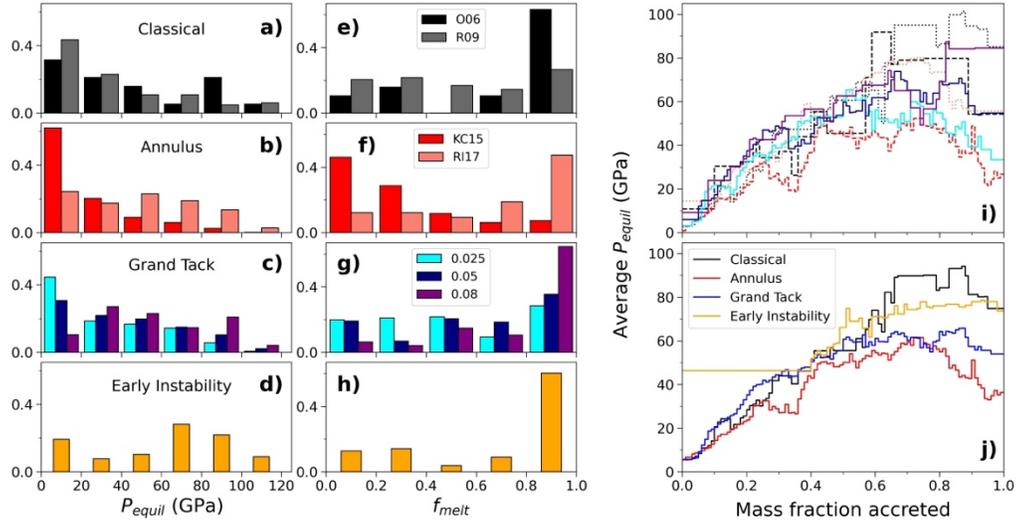

**Fig. 3.** $P_{equil}$ and $f_{melt}$ determined from the melt-scaling law of Nakajima et al. (2021) for embryo impacts. a–d) Normalized histograms of $P_{equil}$ and e–h) $f_{melt}$ for each scenario. $P_{equil}$ and $f_{melt}$ are binned in increments of 20 GPa and 0.2, respectively. i–j) $P_{equil}$ averaged as a function of mass fraction accreted across all Earth analogs from each simulation type excluding the equilibration of planetesimals at 1 GPa to highlight the effects of embryo impacts. Simulation suites are shown separately in i) and grouped by dynamical type in j). The colors of lines in i) correspond to the bar colors in a–h). Abbreviations are as follows: O06 (O'Brien et al., 2006), R09 (Raymond et al., 2009), KC15 (Kaib and Cowan, 2015), RI17 (Raymond and Izidoro, 2017a), and 0.025, 0.05, and 0.08 represent the initial embryo masses, in Earth masses, of Grand Tack simulations from Jacobson and Morbidelli (2014).

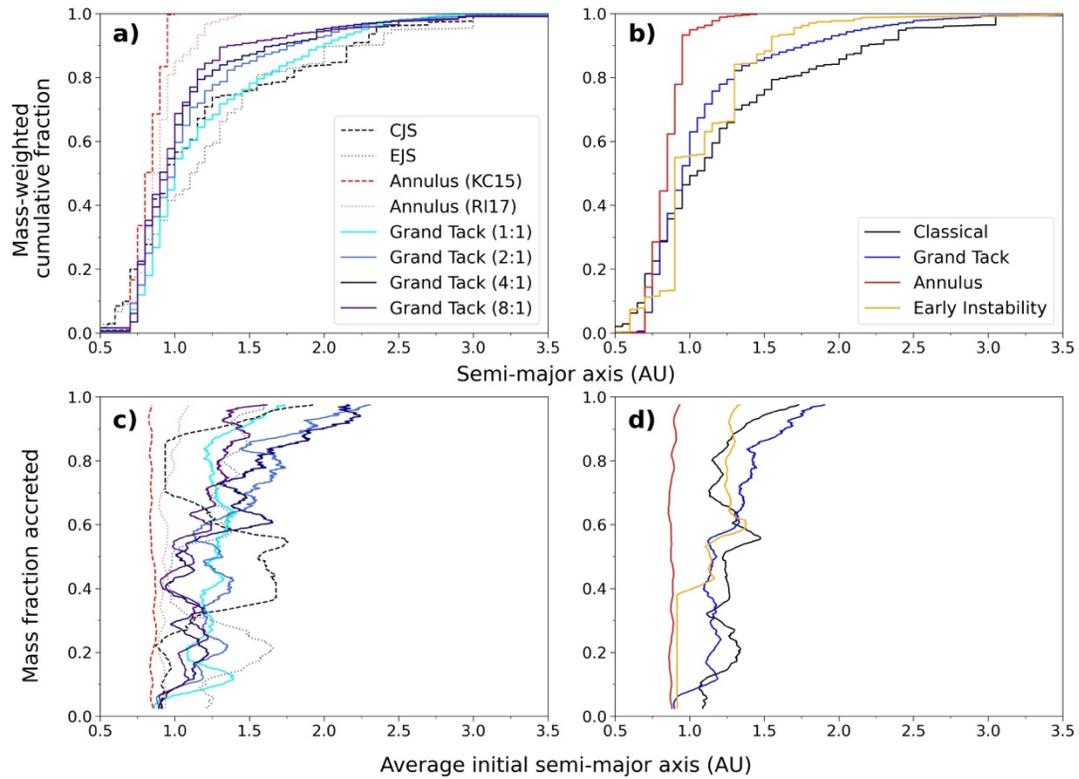

**Fig. 4.** Feeding zones of Earth analogs. a–b) Mass-weighted cumulative distribution of where Earth-forming material originates. c–d) Moving average of the initial semi-major axis of Earth-forming material over the course of Earth's accretion. The initial semi-major axes of Earth accreting material are averaged over windows of 0.05 in mass fraction accreted. Simulation suites are shown separately in a) and c) and are grouped by dynamical type in b) and d). Abbreviations are as follows: CJS (Circular Jupiter and Saturn), EJS (Eccentric Jupiter and Saturn), KC15 (Kaib and Cowan, 2015), RI17 (Raymond and Izidoro, 2017a), and 1:1, 2:1, 4:1, and 8:1 represent the initial embryo to planetesimal mass ratios of Grand Tack simulations from Jacobson and Morbidelli (2014).

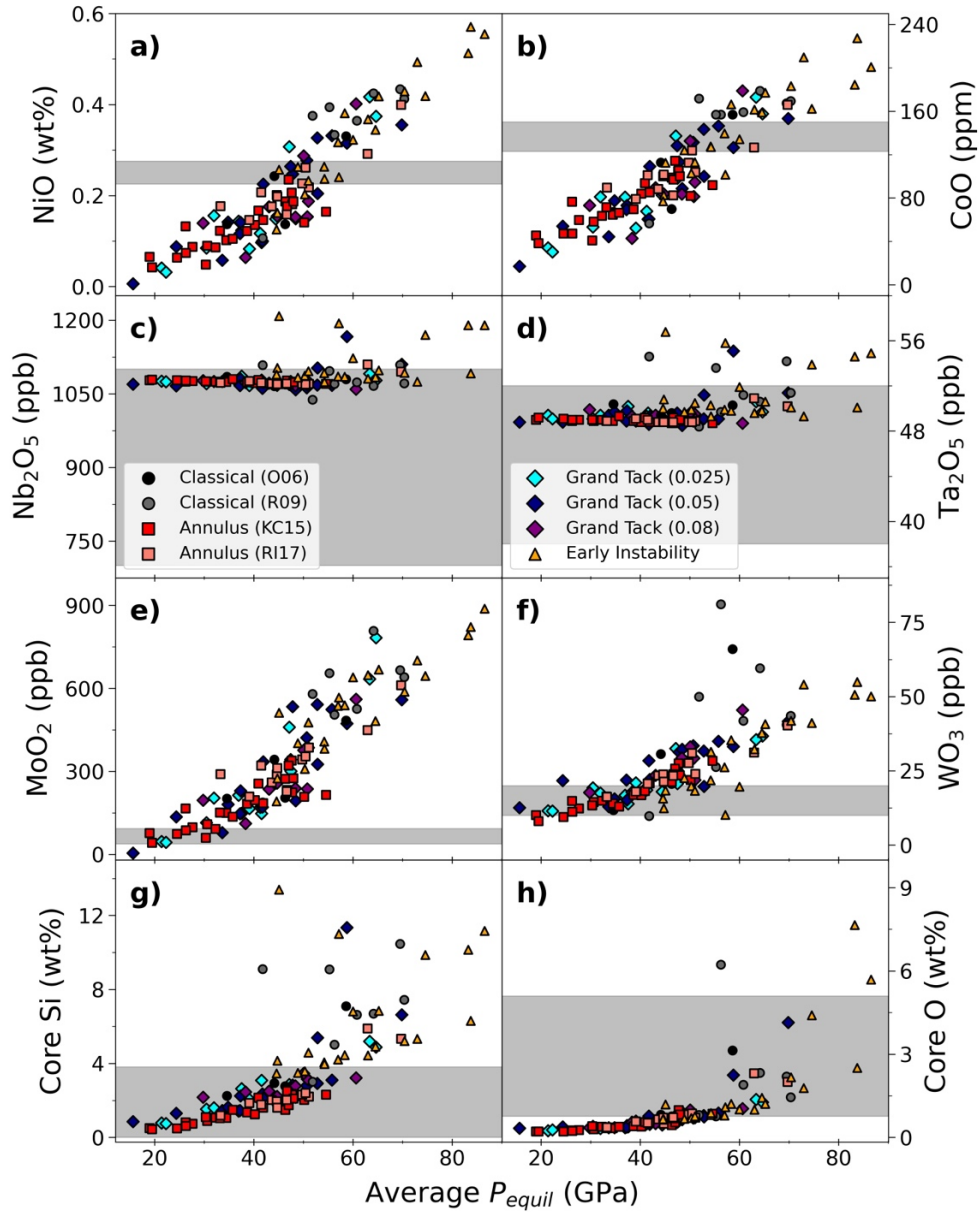

**Fig. 5.** Core and mantle compositions of individual Earth analogs. Data points are core and mantle compositions of individual Earth analogs plotted as a function of $P_{equil}$ averaged over the course of accretion. a–f) Mantle and g–h) core concentrations of investigated elements when model parameters are set to their reference values. Gray shaded regions represent estimated ranges of core and mantle compositions encompassed by Hirose et al. (2021) for the core and from both McDonough and Sun (1995) and Palme and O'Neill (2013) for the mantle. Abbreviations are as follows: O06 (O'Brien et al., 2006), R09 (Raymond et al., 2009), KC15 (Kaib and Cowan, 2015), RI17 (Raymond and Izidoro, 2017a), and 0.025, 0.05, and 0.08 represent the initial embryo masses, in Earth masses, of Grand Tack simulations from Jacobson and Morbidelli (2014).

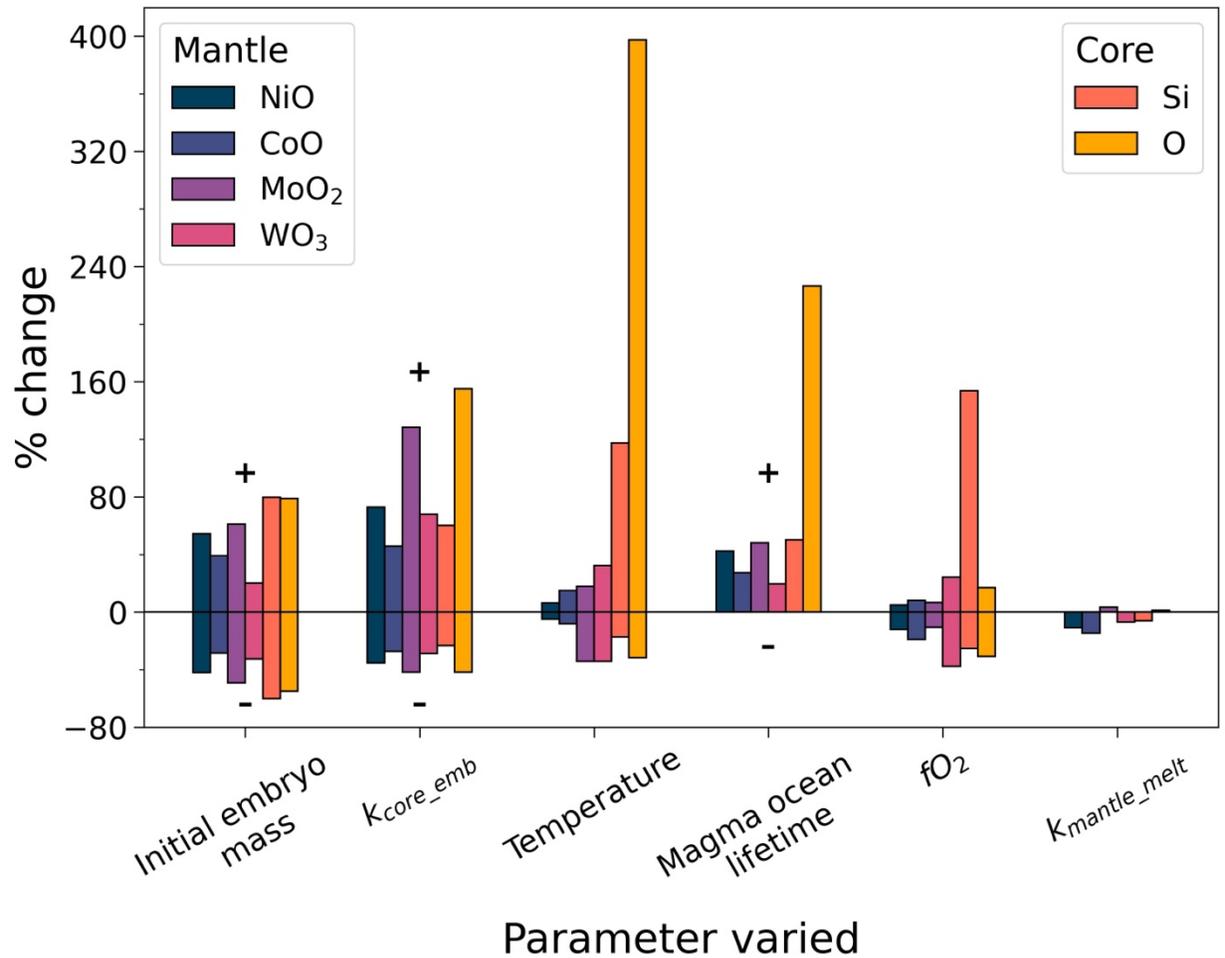

**Fig. 6.** Effects of varying select parameters on core and mantle compositions of Earth analogs. Percent change shows the effect of varying each parameter from the reference value within the ranges specified in Table 2 (expressed as new value minus reference divided by reference). For each set of parameters, the compositions of all Earth analogs are averaged and compared to the average core and mantle compositions from the reference model. Signs ("+" and "–") indicate the direction the parameter is varied to result in the percent change shown. Parameters without signs are those with contrasting effects depending on the element of interest. See Table 2 for the sensitivities of all elements to all parameters.

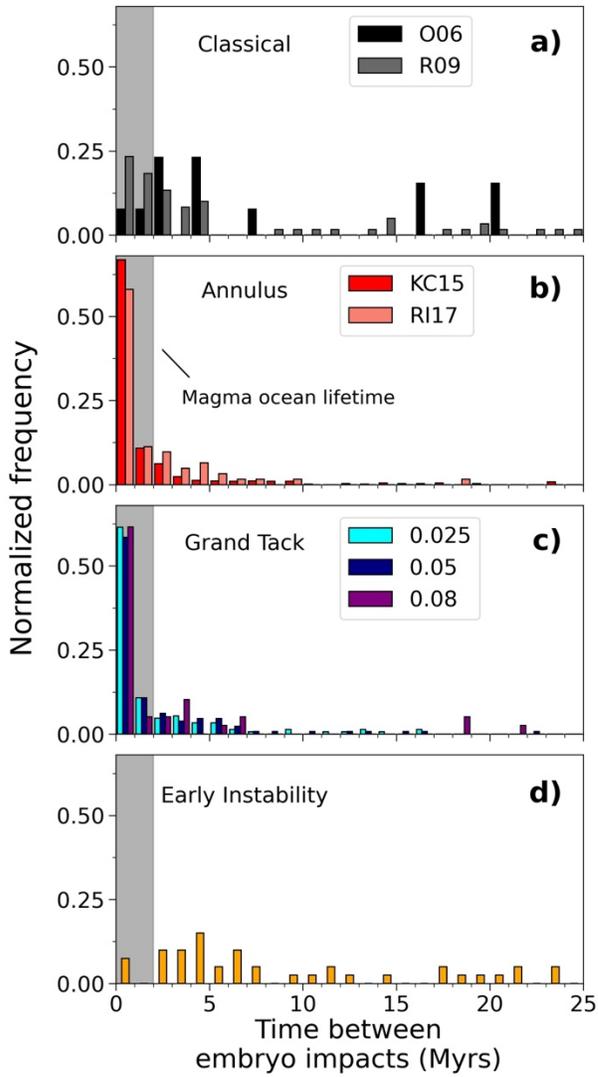

**Fig. 7.** Comparison of a typical magma ocean lifetime with the time between embryo impacts. Normalized histograms of the time between embryo impacts in each simulation suite are shown. Gray shaded regions represent a typical magma ocean lifetime of 2 Myrs (Lichtenberg et al., 2021). Abbreviations are as follows: O06 (O'Brien et al., 2006), R09 (Raymond et al., 2009), KC15 (Kaib and Cowan, 2015), RI17 (Raymond and Izidoro, 2017a), and 0.025, 0.05, and 0.08 represent the initial embryo masses of Grand Tack simulations in Earth masses from Jacobson and Morbidelli (2014).

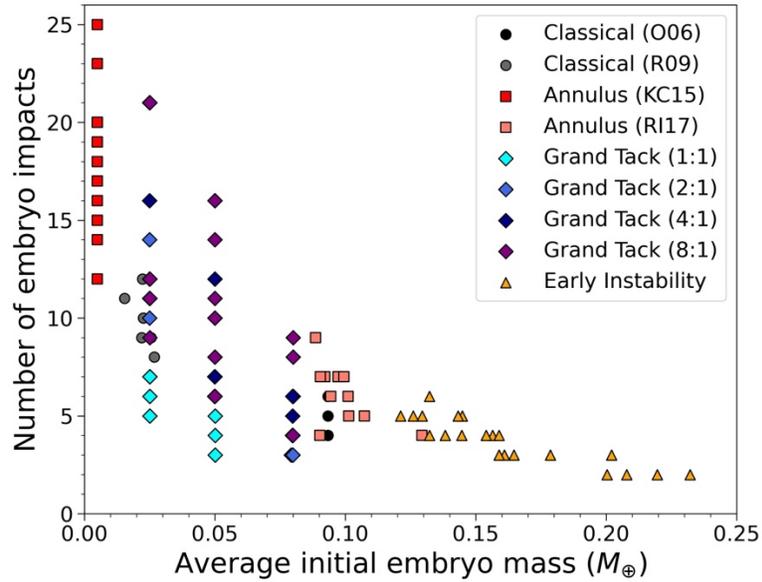

**Fig. 8.** Relationship between the number of embryo impacts experienced by Earth analogs and the average initial embryo mass in each simulation. Average initial embryo masses were calculated for each Earth analog by averaging the initial masses of embryos that eventually collided with the proto-Earth. Abbreviations are as follows: O06 (O'Brien et al., 2006), R09 (Raymond et al., 2009), KC15 (Kaib and Cowan, 2015), RI17 (Raymond and Izidoro, 2017a), and 1:1, 2:1, 4:1, and 8:1 represent the initial embryo to planetesimal mass ratios of Grand Tack simulations from Jacobson and Morbidelli (2014).

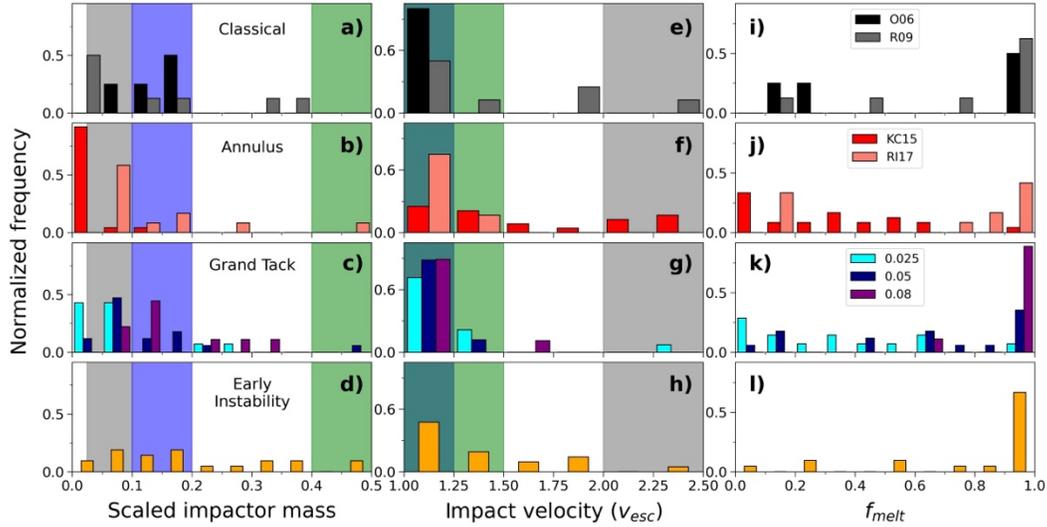

**Fig. 9.** Characteristics of the last embryo impact in each simulation. a–d) Scaled impactor mass ($M_{imp}/(M_{imp} + M_{tar})$), e–h) impact velocity, and i–l) $f_{melt}$. Shaded regions show the a–d) scaled impactor masses and e–h) impact velocities of different moon-forming impactors from the rapidly rotating Earth (gray) (Ćuk and Stewart, 2012), canonical hit-and-run onto a solid or liquid mantle (blue) (Canup and Asphaug, 2001; Hosono et al., 2019; Reufer et al., 2012), and equal sized impactor (green) (Canup, 2012) scenarios. The blue-green regions in e–h are where the canonical hit-and-run and equal sized impactors overlap. Scaled impactor masses, impact velocities, and $f_{melt}$ are binned in increments of 0.05 Earth masses and 0.25, and 0.1, respectively. Abbreviations are as follows: O06 (O'Brien et al., 2006), R09 (Raymond et al., 2009), KC15 (Kaib and Cowan, 2015), RI17 (Raymond and Izidoro, 2017a), and 0.025, 0.05, and 0.08 represent the initial embryo masses, in Earth masses, of Grand Tack simulations from Jacobson and Morbidelli (2014).

**Table 1**

Simulations and characteristics of Earth analogs used in this study. All information in the table is deter

| Simulation type | Earth analogs | Simulations | Disk limits (AU) | Disk surface density $\alpha$[1] | |
|---|---|---|---|---|---|
| Classical (CEJS-O06) | 4 | 8 | 0.3–4.0 | −3/2 | |
| Classical (CEJS-R09) | 8 | 40 | 0.5–4.5 | −3/2 | |
| CJS | 5 | 20 | − | − | |
| EJS | 7 | 28 | − | − | |
| Grand Tack (GT) | 40 | 142 | 0.7–3.0, 6–13[4] | −3/2 | |
|     GT 1:1 | 9 | 34 | | | − |
|     GT 2:1 | 9 | 30 | | | − |
|     GT 4:1 | 8 | 30 | | | − |
|     GT 8:1 | 14 | 48 | | | − |
|     GT-0.025 | 14 | 42 | | | − |
|     GT-0.05 | 17 | 52 | | | − |
|     GT-0.08 | 9 | 48 | | | − |
| Annulus (ANN-KC15) | 24 | 50 | 0.7–1.0 | 0 | |
| Annulus (ANN-RI17) | 12 | 60 | 0.7–1.5 | −1 | |
| Early Instability (EI) | 21 | 133 | 0.48–4.0 | −3/2 | |

[1]Disk surface density is defined as defined as $\Sigma = \Sigma_0 r^{-\alpha}$ where, alpha is the value shown in the column.

[2]Embryo-to-planetesimal mass ratio is the total mass of embryos to the total mass of planetesimals.

[3]Avg $t_{90}$ is the time it takes Earth analogs, on average, to reach 90% of their final mass.

[4]Grand Tack simulations contain mass distributed from 0.7–3.0 AU and planetesimals beyond 6 AU. No

| Embryo mass ($M_{Earth}$) | # embryos | Planetesimal mass ($M_{Earth}$) | # planetesimals |
|---|---|---|---|
| 0.0933 | 25 | 0.0025 | 1000 |
| 0.005–0.15 | 85–90 | 0.0025 | 1000–2000 |
| 0.005–0.15 | – | 0.0025 | – |
| 0.005–0.15 | – | 0.0025 | – |
| 0.025–0.08 | 29–214 | 0.000138–0.0025 | |
| 0.025–0.08 | 28–74 | 0.000138–0.0025 | 727–7209 |
| 0.025–0.08 | 42–131 | 0.000138–0.0025 | 5850 |
| 0.025–0.08 | 55–170 | 0.000138–0.0025 | 2125–4346 |
| 0.025–0.08 | 68–213 | 0.000138–0.0025 | 2250 |
| 0.025 | 74–213 | 0.000138–0.0025 | 727–2250 |
| 0.05 | 37–107 | 0.000138–0.0025 | 727–2250 |
| 0.08 | 28–68 | 0.000138–0.0025 | 2250–7209 |
| 0.005 | 400 | – | – |
| 0.05–0.15 | 20 | 0.000167 | 3000 |
| 0.02–0.48 | 23–25 | 0.0025 | 954–1000 |

mass is initially placed between 3–6 AU.

and EJS simulations and GT simulations are split by initial

| **Embryo:planetesimal mass ratio[2]** | | **Avg $t_{90}$ (Myr)[3]** |
|---|---|---|
| 1:1 | | 53.35 (19.94) |
| 1:1 | | 88.86 (41.98) |
| 1:1 | | 74.33 (3.72) |
| 1:1 | | 77.27 (53.30) |
| 1:1–8:1 | | 23.80 (17.98) |
| | 1:1 | 22.38 (4.27) |
| | 2:1 | 18.70 (4.17) |
| | 4:1 | 31.43 (32.38) |
| | 8:1 | 23.63 (15.82) |
| | 1: 1–8:1 | 26.09 (9.46) |
| | 1: 1–8:1 | 24.39 (23.61) |
| | 1: 1–8:1 | 19.13 (14.65) |
| Inf | | 29.09 (21.17) |
| 4:1 | | 10.18 (4.75) |
| 0.90:1–0.97:1 | | 55.71 (30.57) |

conditions to show the effects of simulation type and initial conditions on Earth analog characterist

| Average mass-weighted semi-major axis (AU) | Percent of material after last giant impact (%) |
| --- | --- |
| 1.27 (0.13) | 7.03 (6.47) |
| 1.27 (0.22) | 1.32 (0.67) |
| 1.23 (0.12) | 1.15 (0.74) |
| 1.29 (0.23) | 3.22 (4.64) |
| 1.15 (0.09) | 8.67 (10.29) |
| 1.23 (0.04) | 17.63 (15.79) |
| 1.16 (0.08) | 10.88 (6.69) |
| 1.13 (0.07) | 5.54 (5.58) |
| 1.10 (0.09) | 3.35 (2.68) |
| 1.15 (0.09) | 6.66 (6.09) |
| 1.14 (0.10) | 9.24 (13.63) |
| 1.19 (0.06) | 10.83 (7.11) |
| 0.85 (0.03) | 0.25 (0.69) |
| 0.92 (0.07) | 6.35 (1.9) |
| 1.16 (0.27) | 0.75 (1.02) |

ics. Numbers in parentheses are standard deviations (±1σ).

**Table 2**
Model parameters, reference values, and the effects of changing each parameter on mantle

| Parameter | Reference | Range | NiO | CoO |
|---|---|---|---|---|
| $P_{equil}$ (GPa) | Melt-scaling[1] | – | – | – |
| Temperature (K) | A11 liquidus[2] | −300 | −4.8% | −8.0% |
| | | +1000 | +6.4% | +15.1% |
| $f_{melt}$ | Melt-scaling[1] | – | – | – |
| $k_{mantle\_melt}$ | 1 | 0.1 | −10.7% | −14.5% |
| | | 1 | – | – |
| $k_{core\_emb}$ | 0.3 | 0.1 | −35.4% | −27.1% |
| | | 1 | +73.1% | +45.9% |
| $P_{equil\_ptsml}$ (GPa) | 1 | 1 | – | – |
| | | 15 | +5.1% | +8.0% |
| $M_{melt\_ptsml}$ (impactor mass) | 5 | 1 | +18.7% | +14.5% |
| | | 10 | −15.4% | −10.9% |
| $k_{core\_ptsml}$ | 0.7 | 0.3 | −6.5% | −3.3% |
| | | 1 | +4.7% | +3.0% |
| $fO_2$ inner (ΔIW) | Match FeO[3] | −3.5 | −11.8% | −19.0% |
| | | −2.5 | +5.0% | +8.1% |
| $fO_2$ outer (ΔIW) | −1.5 | – | – | – |
| $fO_2$ step (AU) | 2 | – | – | – |
| Initial embryo mass[4] | – | ANN-KC15 | −41.8% | −28.4% |
| | | EI | +54.5% | +39.2% |
| Magma ocean lifetime | Instant crystallization | Instant crystallization | – | – |
| | | Infinite MO | +42.4% | +27.5% |

[1]Parameters derived from the melt-scaling law of Nakajima et al. (2021).

[2]Equilibration along the liquidus of Andrault et al. (2011).

[3]Inner $fO_2$ was set for each simulation based on the average FeO of Earth analogs (Fig. S5).

[4]Endmember simulations are those with the lowest and highest average $P_{equil}$ in Fig. 5. Comp

and core compositions of Earth analogs. Changes in composition are calculated from the aver

| Nb$_2$O$_5$ | Ta$_2$O$_5$ | MoO$_2$ | WO$_3$ | Core Si | Core O |
|---|---|---|---|---|---|
| — | | | | — | |
| −0.2% | −0.1% | +18.1% | +32.3% | −17.2% | −31.6% |
| +4.2% | +5.4% | −34.2% | −34.2% | +117.5% | +397.7% |
| — | | — | | | |
| +0.2% | +0.2% | +3.5% | −6.8% | −5.8% | +1.3% |
| — | | | | | |
| −0.0% | −0.2% | −41.6% | −28.8% | −23.2% | −41.5% |
| +1.1% | +0.5% | +128.4% | +68.0% | +60.4% | +155.2% |
| — | | | | | |
| −0.0% | −0.0% | +2.3% | −0.5% | +0.1% | −0.4% |
| −0.0% | −0.0% | +19.8% | +6.2% | +0.1% | −0.1% |
| +0.0% | +0.0% | −16.2% | −2.4% | −0.0% | +0.1% |
| +0.0% | +0.1% | −9.5% | −1.6% | −4.5% | −7.7% |
| -0.0% | -0.0% | +7.0% | +2.4% | +3.2% | +6.0% |
| −3.9% | +10.7% | −10.5% | −37.6% | +153.9% | −30.6% |
| +1.2% | −2.2% | +6.6% | +24.4% | −25.1% | +17.1% |
| — | | — | — | — | — |
| — | | — | — | — | — |
| −0.8% | −1.8% | −49.0% | −32.4% | −60.0% | −54.8% |
| +2.8% | +2.9% | +61.1% | +20.3% | +79.8% | +79.0% |
| — | — | — | — | — | — |
| +3.4% | +3.4% | +48.3% | +19.6% | +50.2% | +226.7% |

positions are averaged over all Earth analogs from each endmember scenario.

raged mantle and core compositions of Earth analogs when one parameter is varied.